# High spectral resolution observations of Uranus' near-IR thermospheric H₂ emission spectrum using the IGRINS spectrograph during the 2018 and 2023 apparitions

Trafton, L. M. & Kaplan, K. F.  The University of Texas at Austin

**Abstract:**  Ground-based near-IR observations have revealed that Uranus' anomalously hot upper atmosphere, detected by Voyager II, has been steadily cooling. The observed $H_3^+$ and $H_2$ emission-line spectra probe Uranus' ionosphere and thermosphere, respectively. Previous observations have shown that the cooling has continued well past the 2007 vernal equinox, when the seasonal solar forcing turned positive, resulting in net heating of the IAU northern hemisphere. Most of them, especially for $H_2$, were obtained at moderate spectral resolution, R ~1000 to 3000, which admits more sky background, with its associated noise, per spectral resolution element relative to spectrographs having higher spectral resolution. We report the first instance of high spectral resolution being used to observe Uranus' fundamental-band rovibrational quadrupole $H_2$ emission spectrum; where the sky background is suppressed and narrow planetary emission lines stand out against the planetary continuum. The IGRINS spectrograph with spectral resolution R ~45,000 was used to observe Uranus in the K-band on Oct 26 & 27, 2018 at the Lowell Discovery Telescope, and on Nov 27, 2023 at Gemini South. These observations reveal rovibrational temperatures of Uranus' thermosphere of 542±25 K and 397±32 K at these two epochs, respectively. The consecutive-nights at elevated temperature observed at the Discovery Telescope suggest that Uranus' near-IR $H_2$ aurora was detected over each of the northern and southern magnetic poles, respectively. The collective IGRINS results support the continued cooling of Uranus' thermosphere through the 2023 apparition, 73% through the spring season.

Corresponding Author:  L. M. Trafton lmt@astro.as.utexas.edu   512 471-1476
McDonald Observatory, Stop C1402:  Austin, TX 78712-1206

## 1. Introduction

The upper atmosphere of Uranus, as for every giant planet, has a temperature that is several hundred degrees hotter than can be accounted for by the absorption of sunlight alone. The cause remains an open question since the Voyager era (Eshleman et al. 1979; Festou & Atreya 1982). For Uranus in particular, the temperature inferred from stellar and solar occultations ($850\pm100$ K) observed by the Voyager 2 Ultraviolet Spectrometer is much higher than can be accounted for by solar extreme ultraviolet heating alone (Herbert et al. 1987; Herbert & Sandel 1994; 1999). Several mechanisms have been proposed to explain the high temperature, but none can account for it fully (Young et al. 2005, Matcheva & Strobel 1999; Schubert et al. 2003; Smith et al. 2007). Uranus stands out because of its high obliquity, which amplifies the seasonal forcing. Furthermore, unlike the other major planets, Uranus has hardly any internal heating. Solar forcing may therefore play a relatively outsized role in the seasonal behavior of Uranus' upper atmosphere. The excessively high temperatures originating in Uranus' ionosphere and thermosphere were confirmed by ground-based observations of the near-infrared emission spectra of $H_3^+$ and $H_2$ (Trafton et al. 1993; 1999).

Continued monitoring the temperature would clarify whether the cooling is seasonal (or the result of a rare exciting event, such as an impact). A reversal in the temperature downtrend before solstice would suggest that the cooling is seasonally driven, so that the orbital (phase lag - between the solar forcing turning positive at equinox and the temperature reversal – would be a measure of the characteristic dynamical time constant of Uranus' thermosphere (Conrath & Pirraglia 1983; Cess & Caldwell 1979). The amount of the phase lag would depend on the nature and details of the processes that transport heat in the thermosphere. But assuming that the seasonal response of Uranus' upper atmosphere behaves like a simple linearized periodic oscillating system that is forced by seasonally changing insolation and damped by thermospheric heat radiating to space, the phase lag cannot extend past a full season (90 deg of orbital phase) (Conrath & Pirraglia 1983). The orbital period of Uranus is 84 yrs, but Uranus is currently near aphelion, so the current season is 23 yrs. A seasonal reversal should thus occur before the IAU northern summer solstice on 4/11/2030. Since radiative transfer is insufficient to explain the observed excess heating, this time constant would be useful for constraining models that explore the role of dynamical processes in driving the seasonal heat transfer in Uranus' upper atmosphere.

Masters et al. (2024) recently proposed an alternative hypothesis for explaining Uranus' long-term cooling; namely, that the cooling of Uranus' thermosphere is the result of a steady, multi-solar cycle, decline in the power of the solar wind incident on the magnetosphere since the Voyager 2 flyby in 1986. The resulting upper atmospheric heating would thus reflect a different process than that of the periodic heating of the space environment over the solar cycle by EUV radiation. We consider this alternative hypothesis further in the Discussion.

Following the detection of $H_3^+$ in Uranus' ionosphere (Trafton et al. 1993), the emission spectra of $H_3^+$ and $H_2$ have been irregularly monitored from multiple observatories, probing both Uranus' ionosphere and thermosphere (Trafton et al. 1999, 2005, 2023; Melin et al. 2011, 2013, 2019, 2020). These are summarized in Figure 1, which shows that Uranus' upper atmosphere has been cooling well past the 2007 Vernal Equinox, when the seasonal forcing became positive, resulting in net hemispheric heating. For most observations, the slit was placed on Uranus' central meridian (CM), with the remainder set across the disk equator. The dashed downtrend lines are weighted least-square linear fits for comparison between $H_3^+$ and $H_2$. Both species show a nearly parallel temperature decline of 260 K since the observations began, with the thermosphere being about 50 K cooler than the ionosphere. This difference occurs because these species effectively probe distinct atmospheric depth regimes: $H_3^+$ is destroyed by a reaction with below the homopause, at O(1 μbar), so only the $H_2$ emission probes the neutral atmosphere (thermosphere), which therefore lies below the homopause, while primarily $H_3^+$ emission probes the ionosphere above it. As of the 2021 apparition, there was no clear sign of a seasonal reversal in temperature.

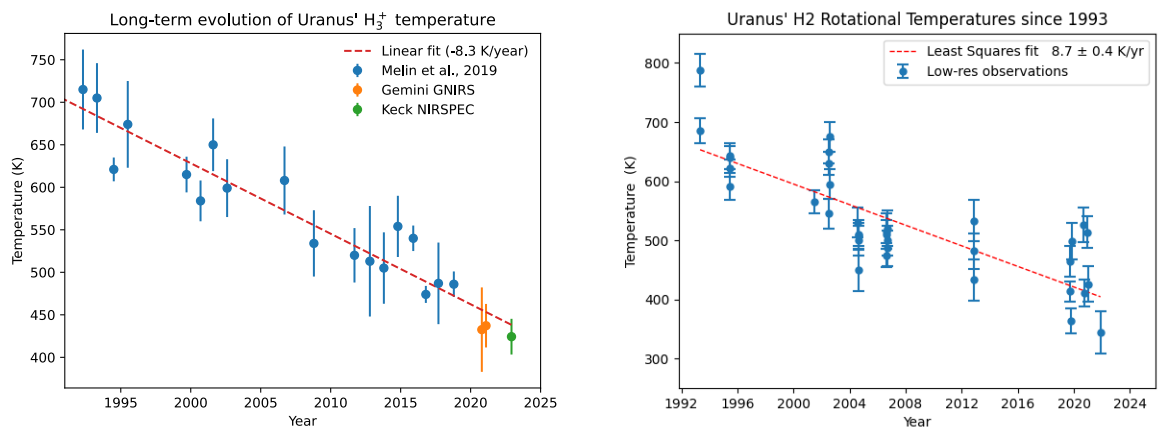

**Fig. 1.** (Left) Long term cooling of Uranus' $H_3^+$ ionosphere – yearly averages (Melin et al, 2019; Trafton et al. 1999, Encrenaz et al. 2000 & 2003. Trafton et al. 2023). (Right) Long term cooling of Uranus' $H_2$ neutral thermosphere. The data are for individual nights during each apparition, (Trafton et al. 1999, 2005; 2023; & in preparation – the recent $H_2$ values are subject to revision.) The dashed lines are linear least squares fits to the data.

## 2. IGRINS observations

The IGRINS (Immersion Grating Infrared Spectrograph) observations extended the $H_2$ monitoring of Uranus through the 2023 apparition. IGRINS is a near-IR cross-dispersed echelle spectrograph with a resolving power R= λ/Δλ = ~45,000 (Yuk et al. 2010, Wang et al 2010, Gully-Santiago et al. 2012, Moon et al. 2012, Han et al. 2012, Park et al. 2014, Oh et al. 2014, Mace et al. 2016, 2018). It uses a silicon immersion echelle grating and two 2Kx2K infrared detectors for simultaneous exposures in the H- and K-band, a fixed slit width, and an infrared slit-viewing camera for rapid acquisition in the K band. It has no cryogenic moving parts. IGRINS is well suited for this program because the fixed slit extends across Uranus' 3."75 disk and the spectral range per exposure in the K-band covers $H_2$ emission lines in both the v=1-0 S-and Q-branches simultaneously. Including lines from each branch allowed us to span a wider range in rovibrational level excitation energy to constrain the rovibrational temperature better. At R=45,000 in the K-band, the sky and planetary continuum background were negligible, and the $H_2$ emission lines were more easily resolved from neighboring telluric water absorption lines.

The previous observations published were taken with longslit spectrographs at various observatories, recently with GNIRS at Gemini North. IGRINS differs by being an echelle spectrograph. Positives for echelle spectrographs are that their higher spectral resolution reduces the continuum background when observing planetary emission lines and helps to separate nearby or blending telluric lines. Negatives are the reduced throughput from the cross-dispersion grating and the associated additional reflections. For IGRINS, the throughput for photons at a given K-band wavelength is $0.06 – 0.08$ (private communication). For GNIRS by comparison, with the blue camera and 111 l/mm grating, the longslit throughput is 0.24 (www.Gemini.edu/Instrument/GNIRS/Sensitivity web page), though we used the 32 l/mm grating, which presumably would alter the GNIRS throughput. An additional negative relative to a longslit spectrograph is the 50% reduction in efficiency when having to nod to sky – as opposed to nodding along the slit – when observing extended objects that extend to more than about a third of the echelle order spacing as projected onto the sky. This is the case for IGRINS since the K-band order spacing at Gemini was 5" ( ~10" at DCT) while Uranus' diameter is 3."8. Further experience observing Uranus with IGRINS and GNIRS should reveal which is better suited for this program.

**Table 1**
Observations and results.

| Observing site | Discovery Channel Telescope | | Gemini South |
|---|---|---|---|
| UT Date obs | Oct 26, 2018 | Oct. 27, 2018 | Nov 27, 2023 |
| UT Start/End | 7:15 – 8:10, 8:26 –10:20* | 9:16-10:04 | 2:51 – 4:45 |
| On+Off Target Exp. time | 2:15 | 0:40 | 1:40 |
| | | Total 2:55 | |
| Mean Sub-Earth E longitude | 207 deg | 8 deg | 312 deg |
| Uranus diameter | 3."73 | | 3."79 |
| Projected slit width | 0."63 | | 0."34 |
| Slit position angle | 79 deg  (slit reversed) | | 271 deg |
| Seeing | 0."9 | 0."6 | 0."5 |
| Uranus sub-Earth latitude | +44 deg  (IAU) | | +63 deg  (IAU) |
| Uranus airmass range | 1.16 -> 1.54 | 1.30 -> 1.48 | 1.510 -> 1.567 |
| Nod amplitude to sky | 30" S | 30" N | ±10" perpendicular to slit |
| Std. Star | HD7215 | | HIP18717 |
| Std. Star airmass | 1.09->1.55 | 1.48 | 1.52 |

*The DCT observation of Uranus on 10/26/18 was interrupted by 16 min to observe the standard star. Exposure time excludes the overhead time taken to nod the telescope between planet and sky.

IGRINS was used to observe Uranus at the Discovery Channel Telescope (DCT) (now the Lowell Discovery Telescope) on Oct 26 & 27, 2018, and at Gemini South on Nov 27, 2023. The position angle of the slit was set along Uranus' CM, aligned at 271 deg at Gemini South, putting IAU Uranus North "Up" on the monitor, but reversed to ~79 deg at the DCT. The DCT telescope was nodded 30" N to the sky in an ABBA pattern with 300 s exposures. The same exposure time was used at Gemini South but the nod to the sky was ±10" perpendicular to the slit. The image scale and projected slit width were different at the two telescopes owing to the unequal focal lengths. Table 1 summarizes the observation parameters

For each observation set, we observed an A0 V star at an airmass similar to that of Uranus to use as a standard empirical calibrator for telluric absorption lines and the absolute flux. The standard stars are listed in Table 1.

The two IGRINS observing runs were allocated two nights each; but at Gemini South, only one night was observed due to scheduling pressures stemming from Gemini South's inoperability during the initial part of Semester 2023B. Consequently, the signal-to-noise ratio (S/N) obtained was less than what was achieved at the Discovery Channel Telescope. The comparison of S/N between Gemini and DCT can be determined from the ratio of the observed line intensity to its uncertainty as listed in Table 2; e.g. S/N=18.5 and 12.1 for Q(1) at DCT vs. Gemini. Therefore, the IGRINS DCT observations, which differ in CM longitudes by ~165 deg, offer a broader longitudinal average of Uranus' temperature. The combined nights spanned 60 degrees of longitude, compared to the 35-degree coverage achieved by IGRINS at Gemini South.

## 3. Data Reduction, flux calibration, and telluric correction

The IGRINS data were reduced using the *IGRINS Pipeline Package (PLP)* v3.0 (Kaplan et al. 2024). The *IGRINS PLP* performs standard data reduction routines including flat fielding, readout pattern removal, cosmic ray removal, flexure correction, combining the individual exposures, rectification and extraction of the individual echelle orders, and determining the wavelength solution from OH sky emission lines. We carried out our data analysis with the Python code *plotspec,* which is designed to analyze 2-D (extended) IGRINS spectra and $H_2$ emission lines. The $H_2$ physical constants and line wavelengths are taken from Roueff et al. (2019).

To empirically calibrate the absolute flux and correct the telluric absorption in Uranus' spectrum, one needs to know the intrinsic spectrum of the standard star and the fraction of the star's light that passes through the slit. We estimated the standard star's intrinsic spectrum by constructing a synthetic spectrum with the software *gollum* (Shanker et al. 2024) from a grid of Phoenix stellar atmosphere model spectra (Husser et al. 2013), using the stellar parameters that implicitly best recreate the telluric-corrected observed standard star's photometric magnitudes and H I line profile shapes. Sources of error were propagated along this process by summing their variances.

The fraction of starlight passing through the slit was estimated by reprojecting the star's PSF measured along the slit length into a 2-D axisymmetric image and then applying a mask in the shape of the IGRINS slit. Since IGRINS observes the H & K bands simultaneously, this procedure was done for the H and K bands separately and the difference in PSF width between the two bands was used to estimate the wavelength-dependent slit loss for the star. As a check on the procedure for determining the fraction of starlight entering the slit at the DCT telescope, we obtained an image of the star, using the slit camera, that was centered on the slit for the spectral exposure, and compared it to another image of the star reflected entirely by the slit. However, this check was not available for the observation at Gemini owing to the strong saturation of the star due to the excellent seeing.

To get the empirical absolute flux calibration and telluric correction, we divided the observed standard star spectrum by the synthetic spectrum. Then we divided by the estimated fraction of starlight through the slit. We then divided our observed Uranus spectrum by this calibration to cancel the instrument response and get our final absolute flux-calibrated and telluric-corrected Uranus spectrum in units of erg cm$^{-2}$ s$^{-1}$ μm$^{-1}$. The statistical uncertainty was propagated by similarly calibrating the variance of the spectrum.

## 4. Results

We detected up to five $H_2$ rovibrational transitions arising from the v=1 vibrational band in our observations of Uranus: 1-0 S(0), 1-0 S(1), 1-0 Q(1), 1-0 Q(2), and 1-0 Q(3). To measure the $H_2$ line intensities, we merged all the echelle orders, interpolated all the $H_2$ lines into position-velocity space, where "velocity" is simply a measure of the spectrum's wavelength via v=c*Δλ/λ, c is the speed of light, and Δλ is the offset from the expected line center. We summed Uranus' flux spectrum along the spatial axis of the slit, which is set along the central meridian to collapse the 2-D spectrum into 1-D. Therefore, Fig. 2 shows spectra summed over Uranus' central meridian. The S/N is too poor to show the spatial profiles. We divided the line fluxes by the associated solid angle Uranus subtends on the IGRINS slit to get the mean intensity in units of erg cm$^{-2}$ s$^{-1}$ (km s$^{-1}$)$^{-1}$ sr$^{-1}$. The solid angle differs by 55% between the two telescopes (Table 1). Nearby pixels in velocity were used to estimate and subtract the background.

The low signal level, and nearby telluric absorption residuals for some of the lines, led to noisy $H_2$ emission line profiles in our Uranus spectra. To mitigate the impact of noise on the line area, crucial for extracting the line intensities, we employed a bright, well-isolated, night sky OH emission line observed in one of the OFF (nodded to night sky) exposures to represent the IGRINS line spread function and similarly interpolated it into position-velocity space. This is because all the OH lines in the K-band have similar line profiles. The IGRINS spectral resolution of R=45,000 gives a FWHM= 0.53 A at the H2 Q-branch Q(1) line (24066 A), while the Doppler FWHM at 400-500 K is 0.24-0.27 A. Combining these Gaussian profiles quadratically results in a 10-11% increase, respectively, in the observed $H_2$ FWHM line width. Given that the full width at half maximum (FWHM) of the OH sky line approximates the instrumental dispersion, and the width of the quadrupole $H_2$ lines, all that is needed for the extraction of the line strength is to scale the OH line area to fit that of the observed emission line profile. The resulting fitted value represents the extracted line intensity.

In practice, we conducted an optimal extraction (Horne 1986) weighting the extraction with the normalized night sky OH line profile and the inverse propagated statistical variance per pixel. Figure 2 shows 1-D profiles for all observed $H_2$ lines and the OH line profiles used for weighting the optimal extraction scaled to the extracted $H_2$ line intensities. The optimal extraction has the advantage of significantly down-weighting noise spikes in the wings of the $H_2$ line profiles, providing more robust extracted $H_2$ line intensities and uncertainties. Since we have already divided the line fluxes by the associated solid angle Uranus subtends on the IGRINS slit, the optimal extraction gets us the mean specific intensity $I_{ul}$ of the $H_2$ lines along the observed CM in units of erg cm$^{-2}$ s$^{-1}$ sr$^{-1}$, which we report along with the $1\sigma$ statistical uncertainty in Table 2.

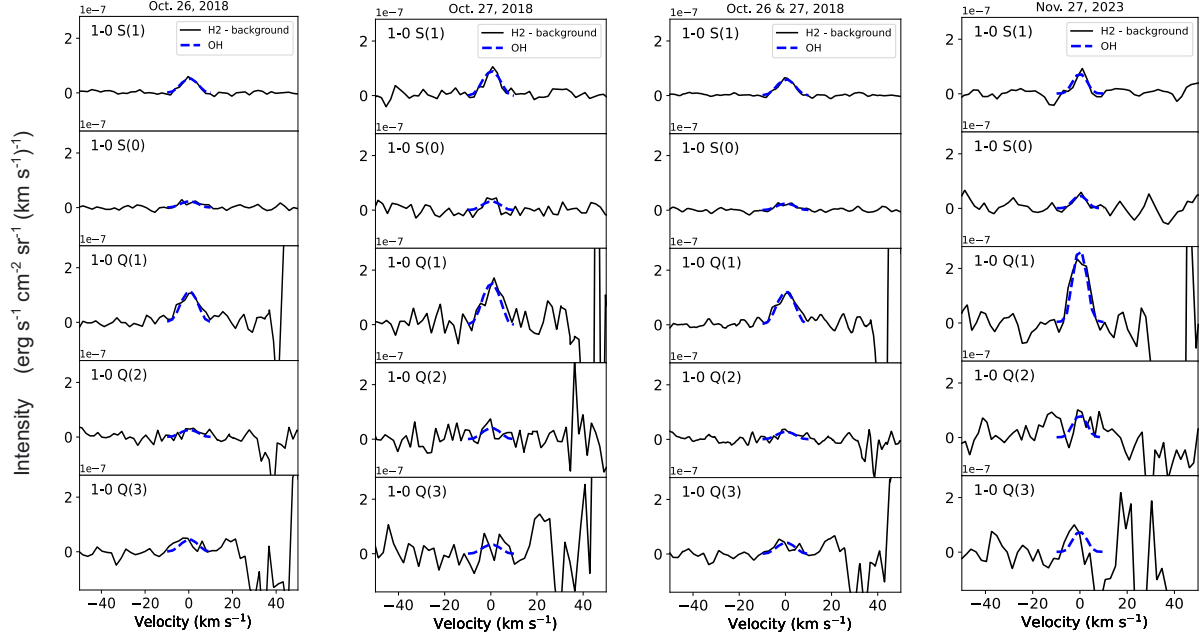

**Fig. 2.** 1-D spectra of all the $H_2$ lines observed in Uranus interpolated into velocity space. The background has been subtracted. The $H_2$ lines are shown in black and the OH line used as the line spread function profile for the optimal spectral extraction weighting is shown as a dashed blue line. The OH spectral line profile shown here is scaled to match the optimally extracted intensity for each line.

**Table 2**
$H_2$ line intensities

| $H_2$ Line | Wavelength | Line Intensity ($I_{ul}$) | | | |
|---|---|---|---|---|---|
| | | Oct. 26, 2018 | Oct. 27, 2018 | Oct 26 & 27, 2018 Combined | Nov 27, 2023 |
| | | (erg cm$^{-2}$ s$^{-1}$ sr$^{-1}$) | | | |
| 1-0 S(1) | 2.1218337 | 4.90e-07 ± 2.48e-08 | 8.31e-07 ± 5.43e-08 | 5.52e-07 ± 2.26e-08 | 5.37e-07 ± 7.13e-08 |
| 1-0 S(0) | 2.2232902 | 2.17e-07 ± 3.36e-08 | 2.94e-07 ± 7.18e-08 | 2.30e-07 ± 3.04e-08 | 3.44e-07 ± 9.07e-08 |
| 1-0 Q(1) | 2.4065919 | 1.06e-06 ± 6.82e-08 | 1.41e-06 ± 1.38e-07 | 1.13e-06 ± 6.11e-08 | 1.98e-06 ± 1.63e-07 |

| 1-0 Q(2) | 2.4134388 | 2.47e-07 ± 8.36e-08 | 3.81e-07 ± 1.73e-07 | 2.63e-07 ± 7.53e-08 | 5.90e-07 ± 1.95e-07 |
| 1-0 Q(3) | 2.4237297 | 4.19e-07 ± 9.44e-08 | 3.06e-07 ± 1.91e-07 | 3.88e-07 ± 8.46e-08 | 5.46e-07 ± 2.00e-07 |

The column density was determined for the $H_2$ vibrational transitions. The $H_2$ emission line intensities are converted into column densities of the upper level of each transition $N_u$ using the equation:

$$N_u = \frac{4\pi I_{ul}}{\Delta E_{ul} hc A_{ul}} \qquad (1)$$

where u and l denote the upper and lower levels of the transition, and $I_{ul}$ is the observed line intensity of the transition averaged over Uranus' CM calculated from the observed line flux and the solid angle of Uranus' CM seen through the slit. $\Delta E_{ul}$ is the difference in energy between the upper and lower levels of the transition, h is Planck's constant, c is the speed of light, and $A_{ul}$ is the radiative transition probability. The values for $\Delta E_{ul}$ and $A_{ul}$ are from Roueff et al. (2019).

Equation(1) is correct only for optically thin radiation. We verified that this is the case for the strongest $H_2$ quadrupole line, Q(1), using the lab-temperature line strength measured by Bragg et al. (1982) of $60.7 \times 10^{-3}$ cm$^{-1}$/ km-Am and correcting the population of the energy levels for the ~400K temperature of Uranus thermosphere. We converted the Bragg et al. line strength units to per molecule/cm$^2$, then divided by the $H_2$ Doppler profile FWHM of 0.0414 cm$^{-1}$ at 400K to estimate the peak optical depth value. We then multiplied by the column density of Uranus' hot thermosphere above 10 μbar, given in Table 3 of Trafton et al. (1999), which yielded an optical depth of $1.47 \times 10^{-3}$. Correcting the v=0, J=1 state population for the elevated temperature gave tau= $1.7 \times 10^{-3}$, which demonstrates that the $H_2$ absorption is optically thin.

Table 3 lists the five observed $H_2$ rovibrational transitions for $H_2$ (v=1), associated wavelengths, upper vibrational and rotational levels $v_u$ and $J_u$, excitation energy above the ground rovibrational level of the upper level of the transition $E_u$, and the upper-level column densities $N_u$ derived from the line intensities using Equation 1. Note that the column density for $H_2$ in a given rotational state is based on the measured line intensity via Eq. 1, without explicit reference to the temperature. However, the observed line intensities and corresponding line column densities refer to the level population distribution of $H_2$ in the v=1 vibrational state, which is sensitive to the temperature.

**Table 3**
$H_2$ line column densities

| $H_2$ Line | Wavelength | $v_u$ | $J_u$ | $E_u$ | Column Density $N_u$ | | | |
|---|---|---|---|---|---|---|---|---|
| | | | | | Oct. 26, 2018 | Oct. 27, 2018 | Oct 26 & 27, 2018 Combined | Nov 27, 2023 |
| | | | | (K) | (cm$^{-2}$) | | | |
| 1-0 S(1) | 2.1218337 | 1 | 3 | 6951.29 | 1.90e+13 ± 9.60e+11 | 3.21e+13 ± 2.10e+12 | 2.14e+13 ± 8.73e+11 | 2.08e+13 ± 2.76e+12 |
| 1-0 S(0) | 2.2232902 | 1 | 2 | 6471.39 | 1.20e+13 ± 1.87e+12 | 1.64e+13 ± 3.99e+12 | 1.28e+13 ± 1.69e+12 | 1.91e+13 ± 5.04e+12 |
| 1-0 Q(1) | 2.4065919 | 1 | 1 | 6148.96 | 3.75e+13 ± 2.42e+12 | 5.01e+13 ± 4.88e+12 | 4.01e+13 ± 2.17e+12 | 7.02e+13 ± 5.78e+12 |
| 1-0 Q(2) | 2.4134388 | 1 | 2 | 6471.39 | 1.24e+13 ± 4.19e+12 | 1.91e+13 ± 8.68e+12 | 1.32e+13 ± 3.77e+12 | 2.95e+13 ± 9.78e+12 |
| 1-0 Q(3) | 2.4237297 | 1 | 3 | 6951.29 | 2.28e+13 ± 5.13e+12 | 1.66e+13 ± 1.04e+13 | 2.11e+13 ± 4.60e+12 | 2.97e+13 ± 1.08e+13 |

Figure 3 shows the excitation diagrams, which are constructed by plotting the natural log of the column density $N_u$ divided by the quantum degeneracy $g_u$ of the upper ro-vibrational levels of all the transitions. An isothermal gas will have spatially constant level populations that follow a linear trend with respect to energy level on these excitation diagrams. The rotational temperature $T_{rot}$ (so-called because it is fit to transitions that arise out of different rotational levels of the same vibrational level) is the inverse of the negative slope of the line. Hotter gas will have shallower negative slopes while cooler gas will have steeper negative slopes. To determine the temperature of the upper atmosphere of Uranus using the observed $H_2$ transitions, we fit a line to the data points plotted on these excitation diagrams using linear-least-squares regression weighted by the propagated inverse variance. While the high signal-to-noise 1-0 S(1) and 1-0 Q(1) transitions dominate the fit, we included the lower signal-to-noise 1-0 S(0), 1-0 Q(2), and 1-0 Q(3) lines in the fit because they help constrain $T_{rot}$ and act as additional checks on the quality of the fit. We note that the low signal-to-noise for 1-0 Q(2) and 1-0 Q(3) is due to nearby deep water telluric absorption features; and we excluded the 1-0 Q(3) transition from the $T_{rot}$ fit for the observations on Oct 27, 2018 due to the significant telluric absorption that affects the red wing of the line.

The fitted IGRINS temperatures $T_{rot}$ are given in Table 4, Figure 3, and over-plotted in Fig. 4 for the two epochs. We find $T_{rot}$ for the Oct 26 & 27, 2018 Discovery Telescope and Nov 27, 2023 Gemini South observations to be $542 \pm 25$ and $397 \pm 32$ K respectively. We also fit both adjacent nights Oct 26 & 27, 2018 separately and get $526 \pm 29$ and $622 \pm 59$ K respectively. The overall trend is consistent with Uranus' cooling but the day-to-day variation is larger than we initially expected. We discuss the derived temperatures in detail in the Discussion section that follows. The total $H_2$ column densities in Table 4 assume thermal equilibrium for the vibrational states. LTE is supported for the emitting layers because the ratio of collisional to radiative times is large compared to unity. The radiative times, given by the Einstein A coefficients for $H_2$, are on the order of $10^{-7}$ s$^{-1}$ (Roueff et al. 2019). By comparison, the per-molecule $H_2$ collision rate near the 1 μ-bar level in the hot thermosphere is on the order of 520 s$^{-1}$, maintaining a Boltzmann population against non-thermal radiative dominance.

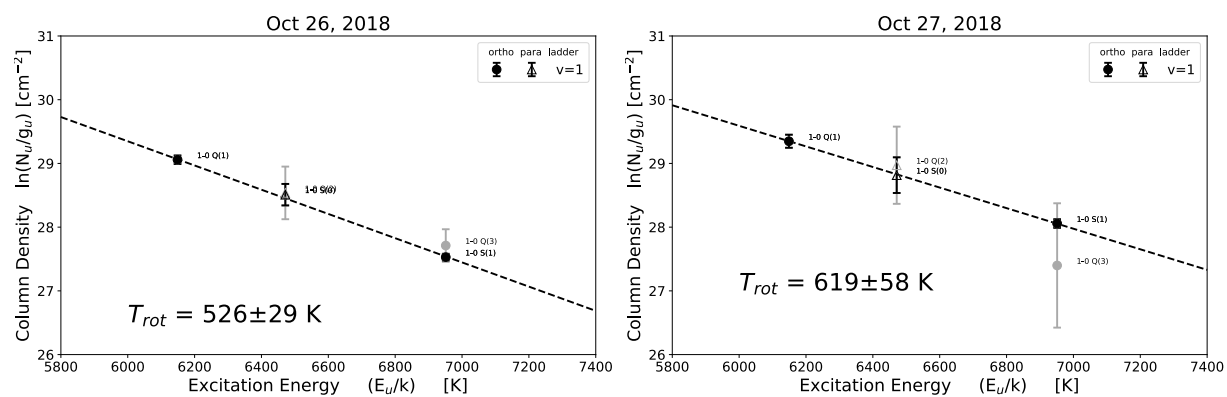

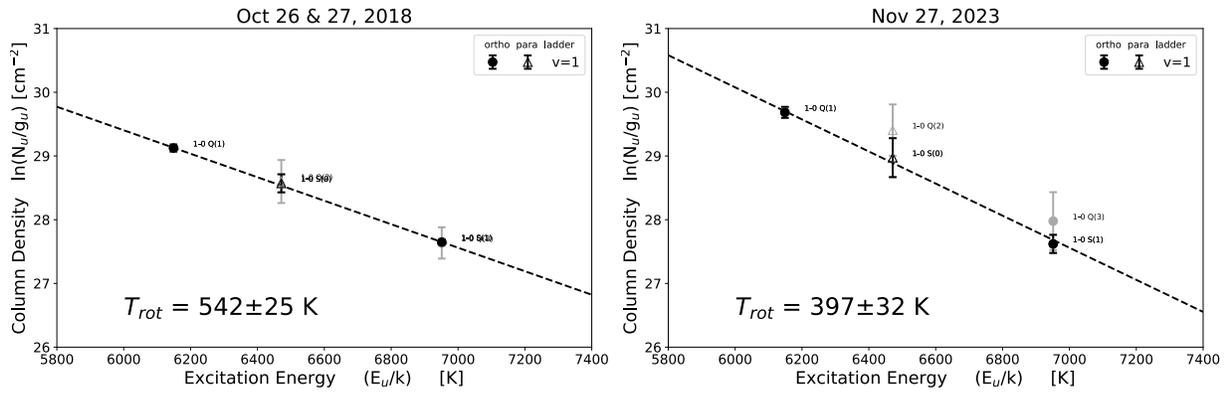

**Fig. 3.** Excitation diagrams for the $H_2$ observed on Uranus on Oct 27 & 28, 2018 (top left and right), the combined Oct 27 & 28, 2018 (bottom left) and Nov 27, 2023 (bottom right). The linear fit to the rotation temperature $T_{rot}$ is shown as the dashed line and value given on each excitation diagram. The $H_2$ 1-0 Q(1), 1-0 S(1), and 1-0 S(0) transitions have the highest signal-to-noise, so they dominate the $T_{rot}$ fits. The other 1-0 Q(2) and 1-0 Q(3) transitions are shown in gray due to their lower signal-to-noise and nearby telluric absorption.

**Table 4**
Uranus $H_2$ rotation temperatures and total column densities

| | Oct. 26, 2018 | Oct. 27, 2018 | Oct 26 & 27, 2018 Combined | Nov 27, 2023 |
|---|---|---|---|---|
| $T_{rot}$ (K) | $526 \pm 29$ | $619 \pm 58$ | $542 \pm 25$ | $397 \pm 32$ |
| $N_{tot}$ (cm$^{-2}$) | $6.51e18\ ^{+6.46e+18}_{-2.93e18}$ | $1.76e18^{+3.32e+18}_{-1.01e18}$ | $5.07e18^{+3.83e+18}_{-1.99e18}$ | $3.97e20\ ^{+1.23e+21}_{-2.73e+20}$ |

Table 4 lists the total $H_2$ column and temperature for Uranus' hot thermosphere extracted on the different dates, assuming LTE. The total column density observed along the line of sight was adjusted for the geometric field of view to correspond to the radial column of hot $H_2$ spanning Uranus' thermosphere. The total $H_2$ column increases by a factor of 78 from the mean of the DCT nights to the night at Gemini. This, along with the large uncertainty in $N_{tot}$, reflects how a small change in temperature, determined by LTE and the few rovibrational levels probed, can drastically affect the inferred total $H_2$ column density. This is because the vast majority of $H_2$ lies in the v=0 low-J levels; i.e., at low excitation energies. At ~450 K, the fractional population of $H_2$ (v=1) is of order 2x10$^{-6}$. Thus, small differences in the temperature measured in the v=1 levels at $E_u/k$ = 6000-7000 K leads to a large change in the overall total column density of $H_2$. Even when the $H_2$ is in LTE, a temperature gradient along the observed column of emitting $H_2$ (such as from a radial variation in the temperature of the thermosphere) cannot be ruled out. This would skew the actual total $H_2$ column density, estimated in Table 4 based just on the v=1 low-J rovibrational levels.

Figure 4 shows the IGRINS rovibrational temperatures added to Fig. 2, extending the monitoring through the 2023 apparition. The outlying single elevated point for 2018 represents an average of elevated temperatures extracted on the two consecutive DCT nights (Table 4). The downtrend line is unchanged from Fig. 2 to facilitate comparison to the added IGRINS data. The excess temperature relative to the down-trending line at the DCT did not occur during the 2023 observation at Gemini South. The 2023 IGRINS temperature is consistent with continued cooling of Uranus' thermosphere.

The error bars of the individual nightly temperatures extracted from the various instruments, including IGRINS, is fairly uniform over the period shown, though it appears somewhat lower for IGRINS. However, the temperature dispersion of the individual exposures relative to the downtrend line is significantly greater, with upward excursions being greater than the downward ones. These peaks might result from localized heating by the low-latitude aurorae due to Uranus' strongly tilted magnetic poles. This suggests that these excursions may represent the random visibility of an auroral pole at the time of the exposure, with the associated partial auroral heating. The greater upward excursions, as for GNIRS at the DCT, may reflect geometries providing greater auroral visibility. Observing runs showing larger ranges of nightly temperatures may have geometries producing a higher "duty cycle" of auroral visibility, where the contributing aurora is visible above Uranus' disk longer. This is covered in more detail in the Discussion section. Further analysis is beyond the scope of this paper.

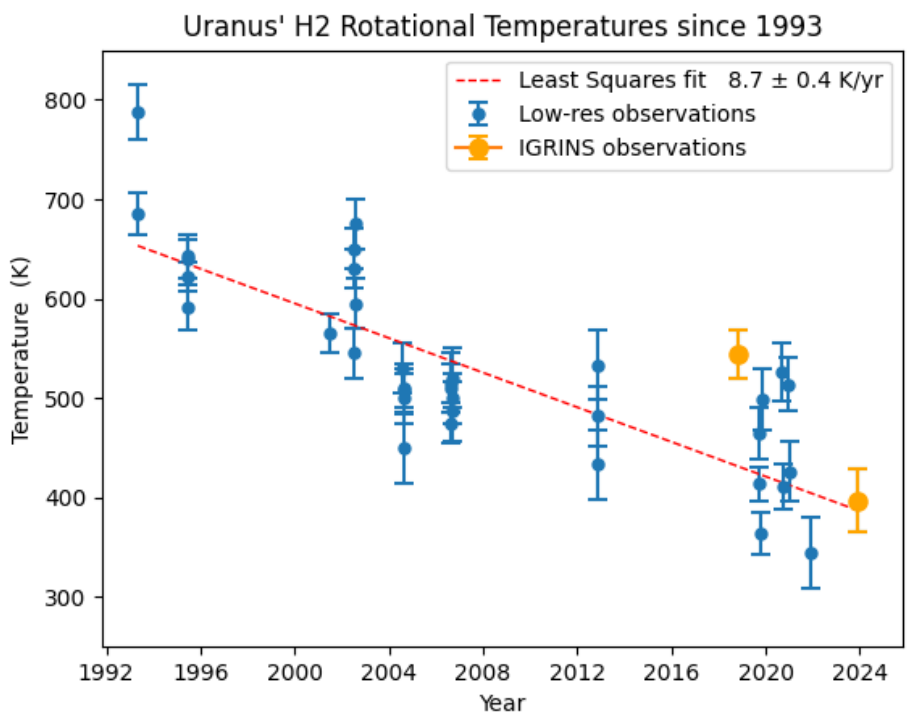

**Fig. 4.** Same as Figure 2 with the two IGRINS observing runs included, indicated by the orange markers. The first combines two nights. The downtrend line is unchanged from Fig. 2 to facilitate comparison with the IGRINS values.

Nevertheless, the temperature downtrend is clear. For most apparitions observed, the dispersed values include the downtrend line, which is fitted without the IGRINS data. This was the case even for the cluster between 2019 and 2023, when the collective dispersion was several times that of a single observation. The systematically lower temperatures during the 2004 apparition is an exception; but passing cirrus clouds and fluctuations in the seeing can affect the flux calibration.

## 5. Discussion

The IGRINS observations at Gemini extend the time-series record to include the 2023 apparition. The temperature extracted from the IGRINS observation on November 2023 is consistent with continued cooling of Uranus' thermosphere.

In presenting their alternative hypothesis, Masters et al. (2024) ruled out the possibility that the temperature downtrend could be seasonal because it continued past the 2007 equinox. However, Conrath & Pirraglia (1983) showed that there should be a seasonal lag in the atmospheric temperature response to changing insolation that is diagnostic of the radiative or dynamical characteristic time constant of the upper atmosphere. To the extent that Uranus' upper atmosphere behaves as a seasonal periodically forced, dissipative, oscillating system, the maximum phase lag would be one season, or 23 years for Uranus near aphelion. Thus, the seasonal hypothesis cannot be ruled out until a continuation of the downtrend past 4/11/2030 is established. To resolve this, continued monitoring is needed.

Moreover, Figure 1 of Masters et al. shows a decline, depicted from the Voyager II epoch, of the solar wind dynamic pressure and kinetic power incident on Uranus' magnetosphere, while the size of magnetosphere increases. The decline is initially in parallel with the decline in the ground-based temperature observed for Uranus' upper atmosphere. However, all three of these processes appear to have leveled off after 2010, while the observed temperature has clearly continued to decline through the 2023 apparition, as of the work reported here. Therefore, both hypotheses remain open, pending further observations.

For a fixed total column density at uniform temperature, an increase in the temperature would normally increase the population of $H_2$ in the vibrationally excited v=1 state, causing the observed emission lines to strengthen with rising temperature, and the thermally-populated rotational population to shift towards higher J levels. Accordingly, of the two adjacent 2018 nights observed at DCT, the observation on Oct. 27 has the higher temperature, line intensity and column density. On the other hand, the vibrationally excited $H_2$ (v=1) emission line fluxes and column densities of the cooler 2023 apparition, which lie much closer to the downtrend line, are only moderately lower than found for the 2018 DCT nights combined (see Tables 2 and 3). The situation is more complicated when a local source of heating, such as an aurora, temporarily appears in the line of sight during planetary rotation.

Aurorae are expected to play a role in heating Uranus' thermosphere and to contribute to its near-IR spectrum (Trafton et al. 1999). Auroral heating can raise the local rovibrational temperature of Uranus' thermosphere. The radial depth over which an aurora can heat the local thermosphere is not clear. But so long as the base of the aurorae lies effectively above the base of Uranus' elevated-temperature thermosphere, the $H_2$ column abundance radiating the aurorae would be less than for the thermospheric emission generally. Also, aurorae tend to be brighter than the thermally emitting thermosphere – locally dominating it. Bright aurora would thus bias the extracted rovibrational temperature and column density.

The differences in the values listed in Tables 2 & 3 between the two 2018 DCT nights show variations higher than their uncertainties, suggesting differences in the $H_2$ line flux and column density with respect to Uranus' longitude. It is also noteworthy that the total $H_2$ column of the combined DCT nights, for which the rovibrational temperature was elevated above the downtrend line, was only 1.4% of the value 5 years later at Gemini, when the thermosphere was cooler. This would imply that the portion of the thermosphere observed during the DCT observations was hotter over a much smaller effective column abundance. Altogether, the outlying results for the 2018 DCT observations might be explained by the contribution of both the northern and southern Uranian aurora on the consecutive nights. This is possible because the longitudes of Uranus' tilted low-latitude northern and southern magnetic poles differ by ~180 deg, which is nearly in agreement with the rotational phase difference between the consecutive DCT nights modulo Uranus' 17.4 hr period. The limited size of an aurora may explain the much smaller column density.

Figure 5 maps Uranus' UV aurorae at both magnetic poles, as observed by the Voyager II UV spectrometer (Broadfoot et al. 1986, Herbert 2009). For reference, the sub-Earth latitude of the DCT observations was IAU +44 deg (IAU North points down on the map). If the geometry and extent of the UV aurorae during the Voyager flyby were representative of the infrared aurore during our observations, there would also be a ~180 deg longitude difference between the two near-IR polar aurorae, corresponding to half of Uranus' rotation period, 8.7 hrs. This duration compares favorably with the time difference between the starts of the two successive DCT observations, modulo Uranus' period, and between their exposure ends, 8.6 & 6.3 hrs, respectively (Table 1).

Based on the geometry of the aurorae in Fig. 5, we estimate that the probability that a random 1 hr exposure with a spectrograph having a narrow slit set along Uranus' CM will continuously observe an aurora is about 30% - for the lower sub-Earth latitudes. The probably of capturing both polar aurorae on consecutive nights given a suitable spacing of the exposure windows is not materially less. The respective DCT observations were 3.1 & 0.48 hrs, (64 & 17 deg of rotation) with the higher flux and temperature observed on the second night (Table 1). If the slit for the latter were centered on the more-compact IAU northern aurora at 235 deg longitude, the fringe of the southern aurora at 55 deg longitude may have been observed on the first night, accounting for its less-elevated brightness and temperature. However, this compact aurora would likely have been missed if instead, the southern aurora were centered on the slit during the second night. Allowing for the exposure duration, the auroral geometry, and Uranus' rotation, the fringes of both aurorae may have been observed during the DCT run, with the slit on the second night being closer to a magnetic pole.

We note that the auroral process is also likely to be variable due to the variable external solar wind forcing (e.g. Masters et al., 2024).

While Lamy 2020 show a spatially sporadic nature of the UV auroral excitation, which results directly from charged particle precipitation, Masters et al. (2024) attribute this largely to the vagaries of the solar wind impinging on Uranus' large magnetosphere. In reality, the magnetic poles are unlikely to have shifted relative to Uranus' body; so the aurorae that are anchored to the poles are unlikely to have shifted either. By contrast, the IR aurorae that are included in our observations may appear less sporadic because, unlike the UV aurorae, they are detected indirectly from the thermal signature that results from the atmospheric heating by the precipitating particles. The IR aurorae are thermalized by kinetic processes and ion chemistry (e.g., Trafton et al. 1999), so they persist according to the local heat capacity and thermospheric winds. Thus, their signature is likely to be more reproducible than for the directly-excited UV aurorae. More observations are needed to characterize the IR aurorae and to update the phase of Uranus' longitudes.

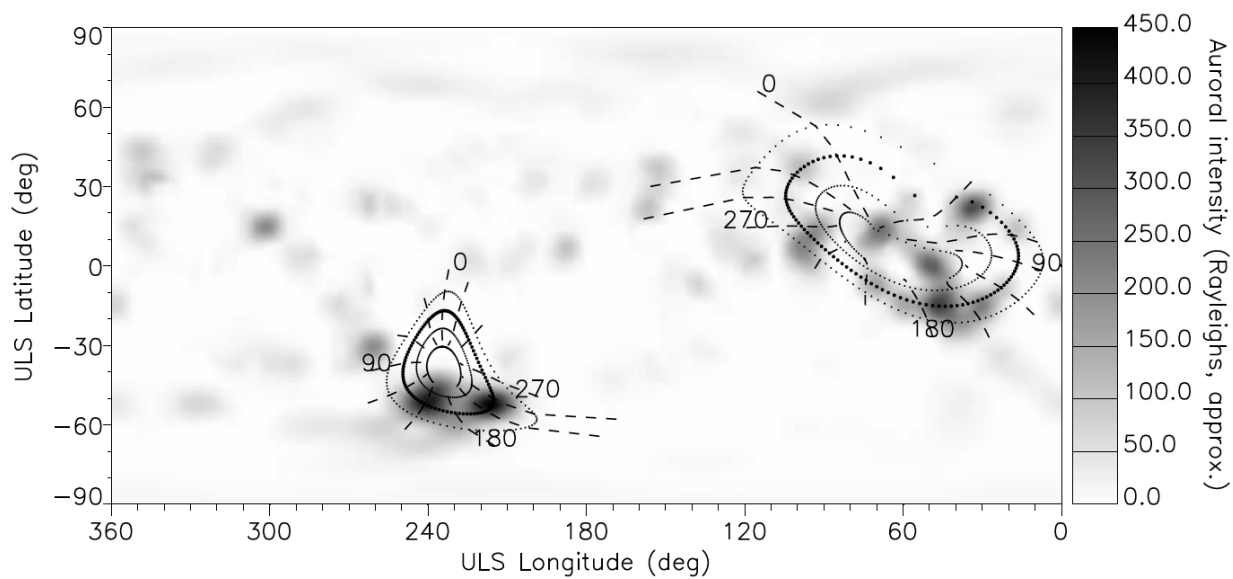

**Fig. 5.** Uranus' UV aurorae observed by the Voyager II UVS (Herbert 2009; Copyright Wiley & Sons; with permission). The negative ULS latitudes refer to IAU North. The sub-Earth latitude for the IGRINS observations was ULS -44 deg. Uranus rotates 20.7 deg per hour. For the observation on Oct 26, Uranus' total rotation was ~64 deg, a third of the longitudinal separation of the auroral poles. On Oct 27, the rotation footprint was ~17 deg, resulting in a combined sub-Earth rotational footprint of ~81 deg. The gaps in longitude between the aurorae are roughly 110 deg & 140 deg.

The JPL Horizons ephemeris tabulates Uranus' longitudes using the rotational period of 17.4 hrs, which was determined during the Voyager II flyby from the rotation of Uranus' magnetosphere. The accumulated drift from using the Voyager era rotational period makes the location of the aurorae in terms of the HORIZONS tabulations undeterminable today. Furthermore, we cannot determine whether the low temperature observed at Gemini corresponds to a non-auroral longitude, to support observing both aurorae at the DCT 5 years earlier, though HORIZONS predicts it. We refer to the Voyager longitudes because the current longitudes are unknown, but longitude differences, such as for the magnetic poles, still have meaning in interpreting observations. Moreover, the JPL HORIZONS planetary ephemeris still uses the Voyager longitude phase, providing a common reference for current researchers.

The episodic visibility of the aurorae every 8.7 hrs should contribute to the temperature dispersion observed over each apparition in Fig. 4, complicating the comparison of the performance between IGRINS and the lower-resolution observations. For that comparison, one should use only the nightly dispersions/uncertainties for each instrument. The nightly temperatures in the three apparitions following the 2018 DCT observation were extracted from Gemini North using GNIRS (Trafton et al. 2020, 2023). The error bars on the extracted line strengths are not appreciably greater than for IGRINS.

## 6. Conclusions

The K-band observations reported here of Uranus' near-IR $H_2$ emission lines taken with the IGRINS spectrograph at the Discovery Telescope in 2018, and at Gemini South during the 2023 apparition, are consistent with a continuation of the long-term temperature downtrend of Uranus' thermosphere, which is now observed to extend through the 2023 apparition. This is 73% into Uranus' IAU northern spring season, despite the rising solar forcing as solstice approaches in 2030. The elevated column abundance of $H_2$ in the v=1 level confirms the elevated rovibrational temperature extracted for each of the two consecutive nights observed at the DCT in 2018. This dual elevation over 26 hrs (178 deg of rotation (modulo 360 deg) vs. ~180 deg of longitudinal separation of the aurorae) suggests the fortuitous detection of each of Uranus' low-latitude aurorae at Uranus' tilted northern and southern magnetic poles - which are expected to be hotter than the thermosphere and are a potential source of the scatter in the temperature time series. Further data that resolve Uranus' near-IR aurorae may improve the precision of the rotational period and reset the drifted phase in longitude. This would improve the planning of observations targeted for auroral studies. We note that solar maximum was predicted to occur in early 2025, which may result in greater auroral activity and improved S/N for the aurorally enhanced temperatures.

## CRediT authorship contribution statement

**Laurence Trafton:** Writing – original draft; Conceptualization; Investigation; Methodology
**Kyle Kaplan:** Writing – review & editing; Methodology; Formal analysis; Software

## Data availability

The IGRINS archive RRISA (https://igrinscontact.github.io) will host the raw and reduced data used in this work following the proprietory period. The data will also be made available upon request.

## Acknowledgements

Based on observations obtained at the international Gemini Observatory, a program of NSF, which is managed by the Association of Universities for Research in Astronomy (AURA) under a cooperative agreement with the U.S. National Science Foundation on behalf of the Gemini Observatory partnership: the U.S. National Science Foundation (United States), National Research Council (Canada), Agencia Nacional de Investigación y Desarrollo (Chile), Ministerio de Ciencia, Tecnología e Innovación (Argentina), Ministério da Ciência, Tecnologia, Inovações e Comunicações (Brazil), and Korea Astronomy and Space Science Institute (Republic of Korea). Program ID GS-2023B-Q-122.

This work used The Immersion Grating Infrared Spectrometer (IGRINS) was developed under a collaboration between the University of Texas at Austin and the Korea Astronomy and Space Science Institute (KASI) with the financial support of the US National Science Foundation under grants AST-1229522, AST-1702267 and AST-1908892, McDonald Observatory of the University of Texas at Austin, the Korean GMT Project of KASI, the Mt. Cuba Astronomical Foundation and Gemini Observatory.

LMT received support from NASA Grant 80-NSSC18K0860 during this research.

We thank Greg Mace for his help with IGRINS in bringing these observations to fruit. We also thank two anonymous reviewers for helpful comments in improving this paper.